\documentclass{appolb}
\usepackage{epsfig}
\begin{document}
\title{Quarkonium dissociation in a PNJL quark plasma%
\thanks{
Poster presented at the EMMI Workshop and XXVI Max-Born Symposium 
``Three Days of Strong Interactions'',
Wroclaw, July 9-11, 2009.
}%
}
\author{Jakub Jankowski
\address{Institute for Theoretical Physics, University of Wroclaw,\\ 
pl. Maksa Borna 9, 50-204 Wroclaw, Poland \\
Institute for Theoretical Physics, University of Amsterdam,\\
Valckenierstraat 65, 1018 XE Amsterdam, The Netherlands}
\and
David Blaschke
\address{Institute for Theoretical Physics, University of Wroclaw,\\ 
pl. Maksa Borna 9, 50-204 Wroclaw, Poland\\
Bogoliubov Laboratory for Theoretical Physics, JINR Dubna, \\
Joliot-Curie Str. 6, 141980 Dubna, Russia }
\and
Hovik Grigorian
\address{
Department of Physics, Yerevan State University, \\
Alex Manoogian str. 1, 375025 Yerevan, Armenia\\
Laboratory for Information Technologies, JINR Dubna, \\
Joliot-Curie Str. 6, 141980 Dubna, Russia }
}
\maketitle
\begin{abstract}
We investigate the Mott effect for heavy quarkonia due to Debye screening 
of the heavy quark potential in a plasma of massless quarks and antiquarks. 
The influence of residual color correlation is investigated by coupling the 
light quark sector to a temporal gauge field driven by the Polyakov loop 
potential. This leads to an increase of the Mott dissociation temperatures 
for quarkonia states which stabilizes in particular the excited states, but
has marginal effect on the ground states. The temperature dependence of
binding energies suggests that the dissciation of the charmonium (bottomonium) 
ground state by thermal activation sets in at temperatures of 200 MeV 
(250 MeV). 
\end{abstract}

%%%%%%%%%%%%%%%%%%%%%
%%%%%%%%%%%%%%%%%%%%%
  
\section{Introduction}
Since the suggestion of $J/\psi$ suppression as a signal of quark-gluon plasma
(QGP) formation by Matsui and Satz \cite{Matsui:1986dk} in 1986 the problem of
quarkonium dissociation in hot and dense strongly interacting matter has played
a key role for QGP diagnostics in relativistic heavy-ion collision experiments.
The original idea was that in a QGP the string tension of the confining 
potential vanishes and the residual one-gluon exchange interaction undergoes 
a Debye screening by the color charges of the plasma. 
When the temperature dependent Debye radius $r_D(T)$ (the inverse of the Debye 
mass $m_D(T)$) becomes shorter than the Bohr radius of the charmonium ground 
state ($J/\psi$) then the Mott effect \cite{Mott:1968zz} (bound state 
dissociation) occurs and the corresponding temperature is 
$T_{\rm Mott}^{J/\psi}$. 
This simple idea grew up to a multifacetted research direction when not only in
the first light ion - nucleus collisions at the CERN NA38 experiment, but also
in proton - nucleus collisions at Fermilab $J/\psi$ suppression has been found
so that there is not only a QGP but also a cold nuclear matter effect on 
charmonium production, see \cite{Rapp:2008tf} for a recent review.

If one wants to explore the question of screening in a plasma more in detail 
then a variety of approaches is available in the literature, from the original 
Debye-H\"{u}ckel approach \cite{Dixit:1989vq} applicable to any vacuum 
potential (for example the Cornell potential),
over the thermodynamic Green functions approach to the ab-initio studies of
heavy-quark potentials in lattice QCD. 
With the obtained medium-dependent potentials one can then study the bound 
state problem  by solving the thermodynamic $T$ - matrix for quarkonia 
\cite{Cabrera:2006wh}, or the equivalent Schr\"{o}dinger-type wave equation 
where medium effects are absorbed in a plasma Hamiltonian 
\cite{Ebeling:1986,Rapp:2008tf}. 

On the other hand one may calculate proper correlators directly from lattice 
QCD and extract from them spectral functions \cite{Asakawa:2000tr}. 
There is an intriguing disagreement between the Mott temperatures deduced from
these spectral functions and those of the potential models: 
%the latter are much smaller than the former! 
From the lattice data for quarkonium correlators one has extracted 
$ T^{\rm Mott}_{J/\psi} \approx 1.9 T_c $  while in 
potential model calculations $ T^{\rm Mott}_{J/\psi} \approx 1.2 T_c$. 
This problem has lead to the discussion of the proper thermodynamical function 
to be used as a potential in the Schr\"odinger equation, see
\cite{Rapp:2008tf,Satz:2005hx} and references therein. 

In this contribution we follow the recently suggested \cite{Jankowski:2009se}
modification of the standard one-loop calculation of the Debye mass in thermal
Quantum Field Theory \cite{LeBellac,Beraudo:2007ky} in the framework of the 
Poyakov-Nambu-Jona-Lasinio model, now widely used for a microscopic 
QCD-motivated description of mesons in quark matter 
\cite{Ratti:2005jh,Hansen:2006ee}. 
We then solve the Schr\"odinger equation for charmonium and bottomonium states 
with the plasma Hamiltonian \cite{Rapp:2008tf} corresponding to the screened 
Cornell potential \cite{Karsch:1987pv} and obtain the Mott dissociation 
temperatures of these states.

%%%%%%%%%%%%%%%%%

\section{Debye-screening in a PNJL quark plasma}

Given the static interaction potential $V(q)$, $ q^{2} = |{\bf{q}}|^{2} $, 
the statically screened potential is given by a resummation of 
one-particle irreducible diagrams ("bubble" resummation = RPA)
\begin{equation}
V_{\rm sc}(q) = {V(q)}/[{1 + F(0;{\bf q})/q^{2}}]~,
\label{Vsc}
\end{equation}
where the longitudinal polarization function 
$ F(0; {\bf q}) = - \Pi_{00}(0; {\bf q}) $ 
in the finite $ T $ case can be calculated within 
thermal field theory as
\begin{equation}
\Pi_{00}(i\omega_{l};{\bf q} ) 
= g^{2} T\sum_{n=-\infty}^{\infty} \int\frac{d^{3}p}{(2\pi)^{3}} 
{\textrm{ Tr}} [\gamma^{0}S_{\Phi}(i\omega_{n}; {\bf p})
\gamma^{0}S_{\Phi}(i\omega_{n}-i\omega_{l}; {\bf p} - {\bf q})]~.
\end{equation}
Here $\omega_{l}=2\pi lT$ are the bosonic and $\omega_{n}=(2n+1)\pi T$
are the fermionic Matsubara frequencies of the imaginary-time formalism.
The symbol ${\textrm{ Tr}}$ stands for traces in color, flavor and Dirac 
spaces.
$S_{\Phi}$ is the propagator of a massless fermion coupled to the homogeneous 
static gluon background field $\varphi_3$. Its inverse is given by 
\cite{Ratti:2005jh,Hansen:2006ee}
\begin{equation}
S^{-1}_{\Phi}( {\bf p}; \omega_{n} ) = 
{\bf \gamma\cdot p} + \gamma_{0}i\omega_{n} -\lambda_{3}\varphi_3~,
\label{coupling}
\end{equation}
where $\varphi_3$ is related to the Polyakov loop variable defined by
\cite{Ratti:2005jh}
$$ \Phi(T) = \frac{1}{3}\rm Tr_c (e^{i\beta\lambda_{3}\varphi_{3}}) 
= \frac{1}{3}(1 + 2\cos(\beta\varphi_3) )~. $$ 
The physics of $\Phi(T)$ is governed by the temperature-dependent Polyakov 
loop potential ${\cal{U}}(\Phi)$, which is fitted to describe the lattice data 
for the pressure of the pure glue system  \cite{Ratti:2005jh}. 
After performing the color-, flavor- and Dirac traces and making the fermionic 
Matsubara summation, we obtain in the static, long wavelength limit 
\begin{eqnarray}
\Pi_{00}( {\bf q} )  
= \frac{2N_cN_f g^2}{\pi^2} \int_{0}^{\infty}dp\, 
p^{2}\frac{\partial f_\Phi}{\partial p} 
%\nonumber
%= - \frac{ 4N_{\rm dof}g^2}{\pi^2} \int_{0}^{\infty}dp\,p f_{\Phi}(p) 
= -2 g^{2}T^{2}I(\Phi) = - m_{D}^2(T) ~,
\label{debyemass}
\end{eqnarray}
where $ m_D(T)$ is the Debye mass, the number of degrees of freedom 
is $N_c=3$, $N_f=2$ and $f_\Phi(p)$ is the quark distribution 
function \cite{Hansen:2006ee}. 
For the discussion of imaginary parts of the polarization function and their
relation to kinetics see, e.g., \cite{Beraudo:2007ky}.
In comparison to the free fermion case \cite{LeBellac,Beraudo:2007ky} the 
coupling to the Polyakov loop variable $\Phi(T)$ gives rise to a modification 
of the Debye mass, given by the integral
\begin{equation}
I(\Phi) = \frac{12}{\pi^2}\int_{0}^{\infty}\,
dx\,x\frac{\Phi(1+2e^{- x})e^{- x}+e^{-3 x}}
{1 + 3\Phi(1 + e^{- x})e^{- x}+e^{-3 x}}.
\end{equation}
The temperature dependence of $\Phi(T)$ is taken from 
%the nonlocal PNJL model of 
Ref.~\cite{Blaschke:2007np}.
In the limit of deconfinement ($\Phi = 1$), the case of a massless
quark gas is obtained ($I(1)=1$), while for confinement ($\Phi = 0$) one finds
that $I(0)=1/9$. 
Taking as the unscreened vacuum potential the one-gluon exchange form  
$ V(q) = -4\pi\alpha/q^2,~\alpha=g^2/(3\pi)$,  the Fourier 
transform of the Debye potential results as statically screened potential,
%
%\begin{equation}
%\label{Vs}
$V_{\rm sc}(q) = -4\pi\alpha/[q^2 + m_D^2(T)]~.$
%\end{equation}
%

%%%%%%%%%%%%%%%%%%%%%%%%%%%%%%%%%%%%
%%%%%%%%%%%%%%%%%%%%%%%%%%%%%%%%%%%%

\section{Schr\"odinger equation and Mott temperatures} 
In order to calculate the temperature dependence of the two-particle energies
$E_{nl}(T)$ for charmonium and bottomonium states in a PNJL quark plasma, we 
solve the Schr\"odinger equation
\begin{equation}
H^{\rm pl}(r;T)\phi_{nl}(r;T) = E_{nl}(T)\phi_{nl}(r;T),
\end{equation}
for the Hamiltonian \cite{Rapp:2008tf}
\begin{equation}
\label{H-pl}
H^{\rm pl}(r;T) = 2m_Q -\alpha m_D(T) - \frac{\nabla^2}{m_Q} + V(r;T),
\end{equation}
with the screened Cornell potential \cite{Karsch:1987pv,Rapp:2008tf}
\begin{equation}
V(r;T) = -\frac{\alpha}{r} e^{-m_D(T) r} 
+ \frac{\sigma}{m_D(T)} [1 - e^{-m_D(T) r}]~,
\label{Potential}
\end{equation}
where parameters are fitted to the vacuum spectroscopy of heavy quarkonia by
$\alpha_s = 0.471 = 3\alpha/4$, $\sigma = 0.192~\textrm{GeV}^2$ and the 
heavy-quark masses $m_c=1.94 $ GeV, $m_b=5.1$ GeV.
Here we use the Debye mass of the previous section with the temperature 
dependence of $\Phi(T)$ taken from a nonlocal PNJL model 
\cite{Blaschke:2007np}.
Note that the Hamiltonian  (\ref{H-pl}) contains a temperature-dependent
shift of the continuum edge due to the Hartree selfenergies of the heavy quarks
in the potential (\ref{Potential}), which results in a definition of the 
dissociation energies as
\begin{equation}
E^{\rm diss}_{nl}(T) := 2m_Q + \frac{\sigma}{m_D} - \alpha m_D - E_{nl}(T)~,
\end{equation}
and of the Mott temperatures 
%corresponding to bound state dissociation 
as $E^{\rm diss}_{nl}(T^{\rm Mott}_{nl})=0$. 

%%%%%%%%%%%%%%%%%%%%%%%%%%%%%%

\begin{figure} [h]
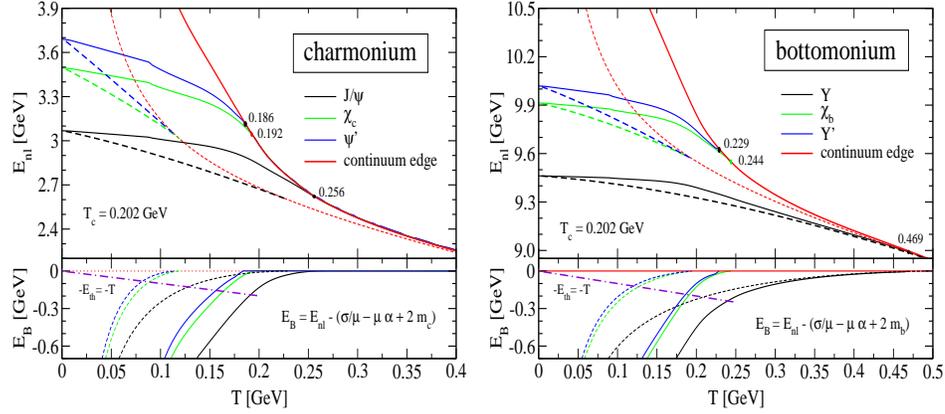

\begin{minipage}{\linewidth}
\begin{minipage}{0.48\linewidth}
%\begin{center}
\includegraphics[width=\textwidth,height=0.9\textwidth]{EB_cc_Tf.eps}
%\end{center}
\end{minipage}
\hspace{0.03cm}
\begin{minipage}{0.48\linewidth}
%\begin{figure}
\includegraphics[width=\textwidth,height=0.9\textwidth]{EB_bb_Tf.eps}
%\end{figure}
\end{minipage}
\end{minipage}

\caption{Upper panels: Temperature dependence of the two-particle energies 
(masses) of the lowest heavy quarkonia states at rest in the medium together 
with the corresponding two-particle continuum edge for charmonia (left) and 
bottomonia (right) with (solid lines) and without (dashed lines) coupling to
the Polyakov-loop potential.
%The coupling to the Polyakov loop potential leads to a suppression of the 
%quark-antiquark excitations in the medium and thus to a stabilization of
%the bound states (solid lines) relative to the case without that coupling 
%(dashed lines). 
Lower panels: Temperature dependence of the binding energies
$E_{\rm B}=-E^{\rm diss}_{nl}$,
compared to the available thermal energy of medium particles 
$E_{\rm th}=T$. For discussion, see text.
%shows that Mott dissociation by thermal activation is
%possible well below the Mott temperatures  estimated from the solution 
%of the Schr\"odinger equation.
} 
\label{Fig:Debye}
\end{figure}

%%%%%%%%%%%%%%%%%%%%%%%%%%%%%%

%The term $-\alpha m_D $ in the Mott condition stems from selfenergy shifts of 
%the heavy quarks in the medium and results in a lowering of the continuum 
%threshold of two-particle scattering states. 
In the upper panels of Fig.~\ref{Fig:Debye} we show the temperature 
dependence of the two-particle energies (masses) of the lowest heavy quarkonia 
states at rest in the medium together with the corresponding two-particle 
continuum edge for charmonia (left) and bottomonia (right).
The coupling to the Polyakov loop potential leads to a suppression of the 
quark-antiquark excitations responsible for the screening of the 
heavy-quark potential in the present model. This entails a stabilization of
the bound states (solid lines) relative to the case without that coupling 
(dashed lines). In the lower panels we show the corresponding binding energies
$E_{\rm B}=-E^{\rm diss}_{nl}$.
Comparison with the available thermal energy of medium particles 
($E_{\rm th}=T$) shows that Mott dissociation by thermal activation 
\cite{Ropke:1988bx} is possible well below the Mott temperatures 
estimated from the solution of the Schr\"odinger equation.
For more detailed discussions of the dissociation kinetics of heavy quarkonia, 
see the recent review \cite{Rapp:2008tf} and references therein.

%%%%%%%%%%%%%%%%%%%%%%%%%%%%%%
%%%%%%%%%%%%%%%%%%%%%%%%%%%%%%

\section{Conclusions}

We have applied the methods of thermal field theory to estimate the effects
of Debye screening on heavy quarkonia bound state formation. In order to
account for residual effects of confining color correlations in the deconfined
phase, we have used the PNJL model in the evaluation of the 
one-loop polarization function. As expected, a stabilization of bound states
in the vicinity of the critical temperature for $T>T_c$ is obtained.
We have solved numerically the Schr\"odinger equation to derive the Mott 
criterion for bound states of the statically screened Cornell potential and 
obtained Mott temperatures in good agreement 
with previous results from nonrelativistic potential models exploiting 
lattice QCD singlet free energies as potentials in the Schr\"odinger equation 
for heavy quarkonia. 
This agreement with previous results (see, e.g., Ref. \cite{Satz:2005hx}) 
includes also higher quarkonia resonances. For these states the 
stabilization effect is most pronounced since their Mott temperatures lie in 
the region of temperatures where the suppression of quark-antiquark excitations
due to the Polyakov-loop potential is the largest.
Due to the lowering of the dissociation energies for heavy quarkonia states, 
their dissociation become possible by thermal activation already well before
the Mott temperatures are reached. According to the present work, even the 
tightly bound bottomonium ground state $\Upsilon$ may get ``boiled away'' at
temperatures $T\sim 250$ MeV, well accessible already at RHIC. 
The detailed discussion of the heavy quarkonia dissociation kinetics and its
relationship to imaginary parts of the heavy-quark potentials 
\cite{Beraudo:2007ky} is beyond the scope of the present contribution.

%%%%%%%%%%%%%%%%%%%%%%%%%
%%%%%%%%%%%%%%%%%%%%%%%%%

\subsection*{Acknowledgments}
The work of DB and JJ has been supported in
part by the Polish Ministry for Science and Higher Education under grant 
No. N N 202 231837. DB acknowledges support from RFFI grant No. 08-02-01003-a.

% ****************************************************************************
% BIBLIOGRAPHY AREA
% ****************************************************************************

\begin{footnotesize}
% IF YOU DO NOT USE BIBTEX, USE THE FOLLOWING SAMPLE SCHEME FOR THE REFERENCES
% ----------------------------------------------------------------------------

% ----------------------------------------------------------------------------

% IF YOU USE BIBTEX,
% - DELETE THE TEXT BETWEEN THE TWO ABOVE DASHED LINES
% - UNCOMMENT THE NEXT TWO LINES AND REPLACE 'Name_Of_Your_BibFile'

%\bibliographystyle{unsrt}
%\bibliography{Name_Of_Your_BibFile}

\begin{thebibliography}{99}
%------- replace following references ;-)

%\cite{Matsui:1986dk}
\bibitem{Matsui:1986dk}
  T.~Matsui and H.~Satz,
  %``J/psi Suppression by Quark-Gluon Plasma Formation,''
  Phys.\ Lett.\  B {\bf 178} 416 (1986).
  %%CITATION = PHLTA,B178,416;%%

%\cite{Mott:1968zz}
\bibitem{Mott:1968zz}
  N.~F.~Mott,
  %``Metal-Insulator Transition,''
  Rev.\ Mod.\ Phys.\  {\bf 40} 677 (1968).
  %%CITATION = RMPHA,40,677;%%

%\cite{Rapp:2008tf}
\bibitem{Rapp:2008tf}
  R.~Rapp, D.~Blaschke and P.~Crochet,
  %``Charmonium and bottomonium production in heavy-ion collisions,''
  arXiv:0807.2470 [hep-ph].% (2008).
  %%CITATION = ARXIV:0807.2470;%%



%\cite{Dixit:1989vq}
\bibitem{Dixit:1989vq}
 V.~V.~Dixit,
  %``CHARGE SCREENING AND SPACE DIMENSION,''
  Mod.\ Phys.\ Lett.\  A {\bf 5} 227 (1990).
  %%CITATION = MPLAE,A5,227;%%


%\cite{Cabrera:2006wh}
\bibitem{Cabrera:2006wh}
D.~Cabrera and R.~Rapp,
  %``T-matrix approach to quarkonium correlation functions in the QGP,''
  Phys.\ Rev.\  D {\bf 76} 114506 (2007);\\
%  [arXiv:hep-ph/0611134].
  %%CITATION = PHRVA,D76,114506;%%\\
%
 C.~Y.~Wong,
  %``Heavy quarkonia in quark gluon plasma,''
  Phys.\ Rev.\  C {\bf 72} 034906 (2005).
%  [arXiv:hep-ph/0408020].
  %%CITATION = PHRVA,C72,034906;%%

%\cite{Ebeling:1986}
\bibitem{Ebeling:1986}
  W.~Ebeling, W.-D.~Kraeft, D.~Kremp, G.~R\"opke,
  {\it Quantum Statistics of Charged Many-Particle Systems},
  Plenum, New York (1986).

%\cite{Asakawa:2000tr}
\bibitem{Asakawa:2000tr}
  M.~Asakawa {\it et al.},% T.~Hatsuda and Y.~Nakahara,
  %``Maximum entropy analysis of the spectral functions in lattice QCD,''
  Prog.\ Part.\ Nucl.\ Phys.\  {\bf 46} 459 (2001);\\
%  [arXiv:hep-lat/0011040].
  %%CITATION = PPNPD,46,459;%%\\
%
A.~Jakovac {\it et al.}, %P.~Petreczky, K.~Petrov and A.~Velytsky,
  %``Quarkonium correlators and spectral functions at zero and finite
  %temperature,''
  Phys.\ Rev.\  D {\bf 75} 014506 (2007).
%  [arXiv:hep-lat/0611017].
  %%CITATION = PHRVA,D75,014506;%%
  
%\cite{Satz:2005hx}
\bibitem{Satz:2005hx}
  H.~Satz,
  %``Colour deconfinement and quarkonium binding,''
  J.\ Phys.\ G {\bf 32} R25 (2006).
%  [arXiv:hep-ph/0512217].
  %%CITATION = JPHGB,G32,R25;%%

%\cite{Jankowski:2009se}
\bibitem{Jankowski:2009se}
  J.~Jankowski and D.~Blaschke,
  %``Quarkonium dissociation in a thermal medium,''
  arXiv:0903.1263 [hep-ph].
  %%CITATION = ARXIV:0903.1263;%%


\bibitem{LeBellac} 
M.~LeBellac, {\it Thermal Field Theory}, Cambridge University Press (1996).

%\cite{Beraudo:2007ky}
\bibitem{Beraudo:2007ky}
 A.~Beraudo, J.~P.~Blaizot and C.~Ratti,
  %``Real and imaginary-time $Q\bar{Q}$ correlators in a thermal medium,''
  Nucl.\ Phys.\  A {\bf 806} 312 (2008).
%  [arXiv:0712.4394 [nucl-th]].
  %%CITATION = NUPHA,A806,312;%%


%\cite{Ratti:2005jh}
\bibitem{Ratti:2005jh}
 C.~Ratti, M.~A.~Thaler and W.~Weise,
  %``Phases of QCD: Lattice thermodynamics and a field theoretical model,''
  Phys.\ Rev.\  D {\bf 73} 014019 (2006).
%  [arXiv:hep-ph/0506234].
  %%CITATION = PHRVA,D73,014019;%%



%\cite{Hansen:2006ee}
\bibitem{Hansen:2006ee}
  H.~Hansen {\it et al.}, 
  %W.~M.~Alberico, A.~Beraudo, A.~Molinari, M.~Nardi and C.~Ratti,
  %``Mesonic correlation functions at finite temperature and density in the
  %Nambu-Jona-Lasinio model with a Polyakov loop,''
  Phys.\ Rev.\  D {\bf 75} 065004 (2007).
%  [arXiv:hep-ph/0609116].
  %%CITATION = PHRVA,D75,065004;%%
  
%\cite{Karsch:1987pv}
\bibitem{Karsch:1987pv}
  F.~Karsch, M.~T.~Mehr and H.~Satz,
  %``Color Screening and Deconfinement for Bound States of Heavy Quarks,''
  Z.\ Phys.\  C {\bf 37}, 617 (1988).
  %%CITATION = ZEPYA,C37,617;%%
%\cite{Blaschke:2007np}

%\cite{Blaschke:2007np}
\bibitem{Blaschke:2007np}
  D.~Blaschke {\it et al.}, %M.~Buballa, A.~E.~Radzhabov and M.~K.~Volkov,
  %``Effects of mesonic correlations in the QCD phase transition,''
  Yad.\ Fiz.\  {\bf 71}, 2012 (2008).
%  [Phys.\ Atom.\ Nucl.\  {\bf 71}, 1981 (2008)]
%  [arXiv:0705.0384 [hep-ph]].
  %%CITATION = PANUE,71,1981;%%

%\cite{Ropke:1988bx}
\bibitem{Ropke:1988bx}
  G.~R\"opke {\it et al.}, %D.~Blaschke and H.~Schulz,
  %``HEAVY QUARK BOUND STATE SUPPRESSION BY MOTT DISSOCIATION AND THERMAL
  %ACTIVATION,''
  Phys.\ Lett.\  B {\bf 202} (1988) 479;
  %%CITATION = PHLTA,B202,479;%%
%
%\cite{Ropke:1988zza}
%\bibitem{Ropke:1988zza}
%  G.~R\"opke, D.~Blaschke and H.~Schulz,
 %``DISSOCIATION KINETICS AND MOMENTUM DEPENDENT J / PSI SUPPRESSION IN A QUARK
  %- GLUON PLASMA,''
  Phys.\ Rev.\  D {\bf 38}, 3589 (1988).
  %%CITATION = PHRVA,D38,3589;%%


\end{thebibliography}
% example of Name_Of_Your_BibFile.bib
% @Article{Turcato:2006ch,
%      author    = "Turcato, M.",
%  collaboration = "ZEUS and H1",
%      title     = "Lepton flavour violation and charmonium physics at HERA",
%      journal   = "Nucl. Phys. Proc. Suppl.",
%      volume    = "162",
%      year      = "2006", 
%      pages     = "283-287",
%      SLACcitation  = "%%CITATION = NUPHZ,162,283;%%"
% }
% 
% @Unpublished{Gogitidze:2007du,
%      author    = "Gogitidze, N.",
%  collaboration = "H1", 
%      title     = "Prompt photons and particle momentum distributions at
%                   HERA", 
%      year      = "2007",
%      note    = "hep-ex/0701033",
%      SLACcitation  = "%%CITATION = HEP-EX 0701033;%%"
% }

\end{footnotesize}

% ****************************************************************************
% END OF BIBLIOGRAPHY AREA
% ****************************************************************************

\end{document}